\documentclass[prr,aps,superscriptaddress,twocolumn,notitlepage,showpacs,nofootinbib]{revtex4-1}
\usepackage{latexsym}
\usepackage{amsmath, dsfont, physics}
\usepackage{amssymb}
\usepackage{graphicx}
\usepackage{caption}
\usepackage{subfigure}
\usepackage{float}
\usepackage{mathrsfs}
\usepackage{color}
\usepackage{txfonts}
\usepackage[justification=centering,
            format=plain]{caption}

\renewcommand{\raggedright}{\leftskip=0pt \rightskip=0pt plus 0cm}

\begin{document}

\title{Engineering bound states in continuum via nonlinearity induced extra dimension}

\author{Qingtian Miao}
\email{qm8@tamu.edu}
\affiliation{Institute for Quantum Science and Engineering, Texas A$\&$M University, College Station, TX 77843, USA}
\affiliation{Department of Physics and Astronomy, Texas A$\&$M University, College Station, TX 77843, USA}
\author{Jayakrishnan M. P. Nair}
\email{jayakrishnan00213@tamu.edu}
\affiliation{Institute for Quantum Science and Engineering, Texas A$\&$M University, College Station, TX 77843, USA}
\affiliation{Department of Physics and Astronomy, Texas A$\&$M University, College Station, TX 77843, USA}
\author{Girish S. Agarwal}
\email{Girish.Agarwal@ag.tamu.edu }
\affiliation{Institute for Quantum Science and Engineering, Texas A$\&$M University, College Station, TX 77843, USA}
\affiliation{Department of Physics and Astronomy, Texas A$\&$M University, College Station, TX 77843, USA}
\affiliation{Department of Biological and Agricultural Engineering, Texas A$\&$M University, College Station, TX 77843, USA}
\date{\today}

\begin{abstract}
Bound states in continuum (BICs) are localized states of a system possessing significantly large life times with applications across various branches of science. In this work, we propose an expedient protocol to engineer BICs which involves the use of Kerr nonlinearities in the system. The generation of BICs is a direct artifact of the nonlinearity and the associated expansion in the dimensionality of the system. In particular, we consider single and two mode anharmonic systems and provide a number of solutions apposite for the creation of BICs. In close vicinity to the BIC, the steady state response of the system is immensely sensitive to perturbations in natural frequencies of the system and we illustrate its propitious sensing potential in the context of experimentally realizable setups for both optical and magnetic nonlinearities.
 
\end{abstract}

\maketitle

\section{Introduction}
The localization of electromagnetic waves has been a subject of intense research over the past few decades \cite{hsu2016bound}. It is well known that the solutions of the Schr\"{o}dinger equation below the continuum threshold possess discrete energies and are square integrable in nature. In contrast, above the continuum threshold, energy eigenvalues are continues and the solutions are unbounded. It has been, however, shown that there exist localized states within the continuum of energies, namely the bound states in continuum. BICs were first proposed in 1929 by von Neumann and Wigner \cite{von1929some} in an electronic system and Stillinger and Herrick later extended it to a two electron wave function \cite{PhysRevA.11.446}. However, a fist experimental observation of BICs came only in 1992 by Capasso $\textit{et al}$, where they demonstrated an electronic bound state in a semiconductor superlattice \cite{capasso1992observation}. 

The emergence of BICs in electromagnetic systems can be explicated by investigating the effective non-Hermitian Hamiltonian ensuing from the Maxwell's equations, resulting in complex resonance frequencies $\omega$. BICs are, in essence, non-radiating solutions of the wave equations, ie., modes of the system with Im($\omega$) approaching zero. In the last decade, BICs have been realized in a multitude of settings involving, for example, electronic \cite{PhysRevB.85.115307, alvarez2015impact, yan2013bound}, acoustic \cite{lyapina2015bound, chen2016mechanical, hein2012trapped} and photonic \cite{PhysRevLett.100.183902} subsystems. In particular, owing to their excellent tunability, photonic systems have emerged as an excellent candidate in recent years with applications including, but not limited to the design of high-Q resonators \cite{zou2015guiding, yu2019photonic, yu2020high}, lasing \cite{kodigala2017lasing, gentry2014dark, midya2018coherent, ha2018directional, wu2020room, azzam2021single, muhammad2021optical, yang2021low}, sensing \cite{ndao2020differentiating, romano2018label, tittl2018imaging, leitis2019angle, zhen2013enabling, sun2016high, wang2018optofluidic, wang2021ultrasensitive, yesilkoy2019ultrasensitive, jahani2021imaging}, filters \cite{foley2014symmetry, doskolovich2019integrated}, etc.  In \cite{romano2018label}, Romano \textit{et al} reported an optical sensor underpinned by BIC for the fine grained estimation of perturbations in a dielectric environment. Another recent work \cite{tittl2018imaging} reported the development of nanophotonic sensor based on high-Q metasurface elements for molecular detection with applications in biological and environmental sensing. Some other recent intriguing research include the enhanced sensing of spontaneous emission \cite{zhen2013enabling}, vortex generation \cite{doeleman2018experimental, wang2020generating}, switches \cite{henkel2021electrically}, efficient higher harmonic generation \cite{koshelev2020dielectric, carletti2018giant} and many more.

In this paper, we propose Kerr nonlinearities as a resource to engineer BICs. Such nonlinearities can be observed in a plentitude physical systems ranging from optical cavities \cite{boyd2020nonlinear} to magnetic systems \cite{PhysRevB.94.224410}, which has been a prime subject of interest, with many exotic effects \cite{PhysRevLett.127.183202, PhysRevLett.126.180401, PhysRevLett.124.213604, PhysRevB.103.224401}. Here, we present a variety of solutions for BICs relevant to single and two mode bosonic systems having a Kerr type of anharmonicity. The resulting BICs are strongly sensitive to perturbations in the system parameters, in particular variations in characteristic detunings which owes its origin to the existence of first and second order poles in the response function. In addition, we discuss a number of experimental platforms germane to our analysis of the nonlinear systems. In particular, we specifically illustrate its sensing capabilities of the two mode anharmonic system in the context of a few experimentally realizable systems.

The manuscript is organized as follows. In section \ref{sec1}, we discuss the well known schemes for the generation of BICs without involving the use of Kerr nonlinearities.  Subsequently, in section \ref{sec2}, we provide a detailed analysis of the protocol to achieve BICs in a single mode system with passive Kerr nonlinearity and the accompanying sensitivity to perturbations in the system. We extend the study into the domain two mode active nonlinear system in section \ref{sec3} and establish its equivalence with the single mode results in Appendix \ref{Appendix:a}. Finally, we conclude our results in section \ref{sec4}.

\section{BIC in a coupled two-mode system} \label{sec1}
We commence our analysis by revisiting the emergence of BICs in a generic two-mode system without any nonlinearities. To this end, we consider a system comprising of modes $a$ and $b$ coupled through a complex parameter $J$ and driven externally at frequency $\omega_d$. The dynamics of the system in the rotating frame of the drive is given by
\begin{align}
\dot{X}=-i\mathcal{H} X+F_{in},
\end{align}
where, $X^T=[$$a$ ${b}$], $F_{in}$ describes the modality of external driving and $\mathcal{H}$ is the effective non-Hermitian Hamiltonian provided by
\begin{align}
\mathcal{H}=\begin{pmatrix}
\Delta_a-i\kappa&J\\
J&\Delta_b-i\gamma
\end{pmatrix}.
\end{align}
Here, $\Delta_i=\omega_i-\omega_d$ where $i\in \{a,b\}$, $\omega_a$ and $\omega_b$ are the characteristic resonance frequencies of the modes $a$ and $b$, and $\kappa$, $\gamma$ denote their respective decay rates. Note that the real and imaginary parts of $J=g-i\Gamma$ represent the coherent and dissipative form of coupling between the modes. The eigenvalues of $\mathcal{H}$ are given by $\lambda_{\pm}=\frac{\Delta_a+\Delta_b}{2}-i\bar{\gamma}\pm\sqrt{(\frac{\Delta_a-\Delta_b}{2}-i\tilde{\gamma})^2+(g-i\Gamma)^2}$, where $\bar{\gamma}=\frac{\kappa+\gamma}{2}$ and $\tilde{\gamma}=\frac{\kappa-\gamma}{2}$. One of the ways to bring to naught the imaginary part of the eigenvalues is to employ engineered gain into the system, that is to make $\kappa=-\gamma$. This in conjunction with the absence of dissipative coupling, \textit{viz}, $\Gamma=0$ and $\Delta_a=\Delta_b$  yield the eigenvalues $\lambda_{\pm}=\Delta_a\pm\sqrt{(g^2-{\gamma}^2)}$. Palpably, the system in the parameter domain $g\geq\gamma$ is earmarked by the observation of real eigenspectra \cite{ruter2010observation}. Note \textit{en passant}, that the system under this parameter choice lends itself to a $PT$-symmetric description of the effective Hamiltonian featuring an exceptional point (EP) in the parameter space at $g=\gamma$. On the other hand, the region $g<\gamma$ affords eigenvalues which form a complex-conjugate pair, wherein the amplitude of the of the modes grows exponentially in time whereas the other one decays. In the context of $PT$-symmetric systems, it is important to notice that EPs, which have found applications in sensing \cite{chen2017exceptional} are functionally analogous to BICs.

The exists another interesting parameter domain, conformable with anti-$PT$ symmetry, i.e., $\{PT,\mathcal{H}\}=0$, that can spawn a BIC, without involving external gain. Such a system necessitates the absence of coherent coupling, that is to say $g=0$, $\kappa=\gamma$ and $\Delta_a=-\Delta_b$, begetting $\lambda_{\pm}=-i\kappa\pm\sqrt{(\Delta_a^2-{\Gamma}^2)}$ which take purely imaginary form when $|\Delta_a|\leq\Gamma$. In contrast, the $|\Delta_a|>\Gamma$ phase leads to decaying solutions with real part of the eigenvalues flanked on either side of the external drive frequency.  Observe that when $\Delta_a= 0$ and as $\Gamma$ approaches $\kappa$, the system entails a BIC, marked by the existence of a vanishing eigenvalue, i.e., $\lambda_+\rightarrow 0$ and thereby eliciting a pole at origin in response to the external drive. The anti-$PT$ symmetric system does not warrant the use of gain, however, it stipulates the use of dissipative coupling, which can be engineered by coupling the subsystems via a common intermediary reservoir \cite{PhysRevB.105.064405, PhysRevLett.123.127202, PhysRevA.100.013812, PhysRevB.103.224401}. 

It makes for a relevant observation that in general, the effective Hamiltonian in Eq. (1) does not yield non-radiating solutions of the Maxwell equations, especially when $J=0$, i.e., when the modes are decoupled. In the following section, we provide a mechanism to engineer BIC in a nonlinear system which does not depend on the underlying symmetries of the system. More importantly, the protocol can be implemented even in the limit where the subsystems are completely decoupled. In fact, the existence of BIC is an inalienable consequence of anharmonicities present in the system and the concomitant magnification of the dimensionality. The mechanism can be extended to two-mode nonlinear systems and we provide a detailed analysis in section \ref{sec3}.

\section{BIC in a single mode Kerr nonlinear system}\label{sec2}
\begin{figure}
 \centering
   \includegraphics[scale=0.46]{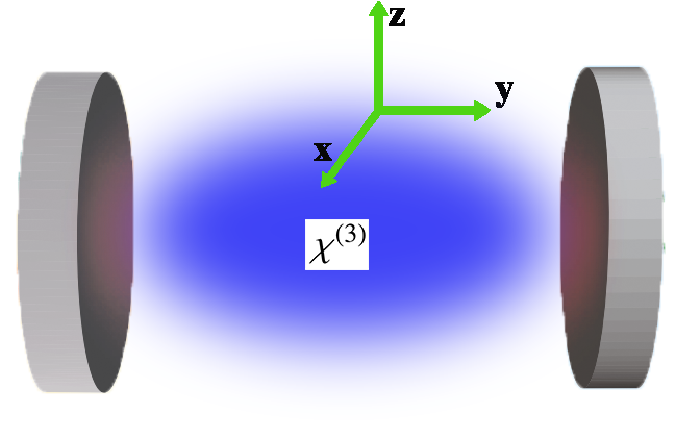}
\caption{A third order Kerr nonlinear medium in an optical cavity.}
\label{fig_0}
\end{figure}
\begin{figure}
 \captionsetup{justification=raggedright,singlelinecheck=false}
 \centering
   \includegraphics[scale=0.46]{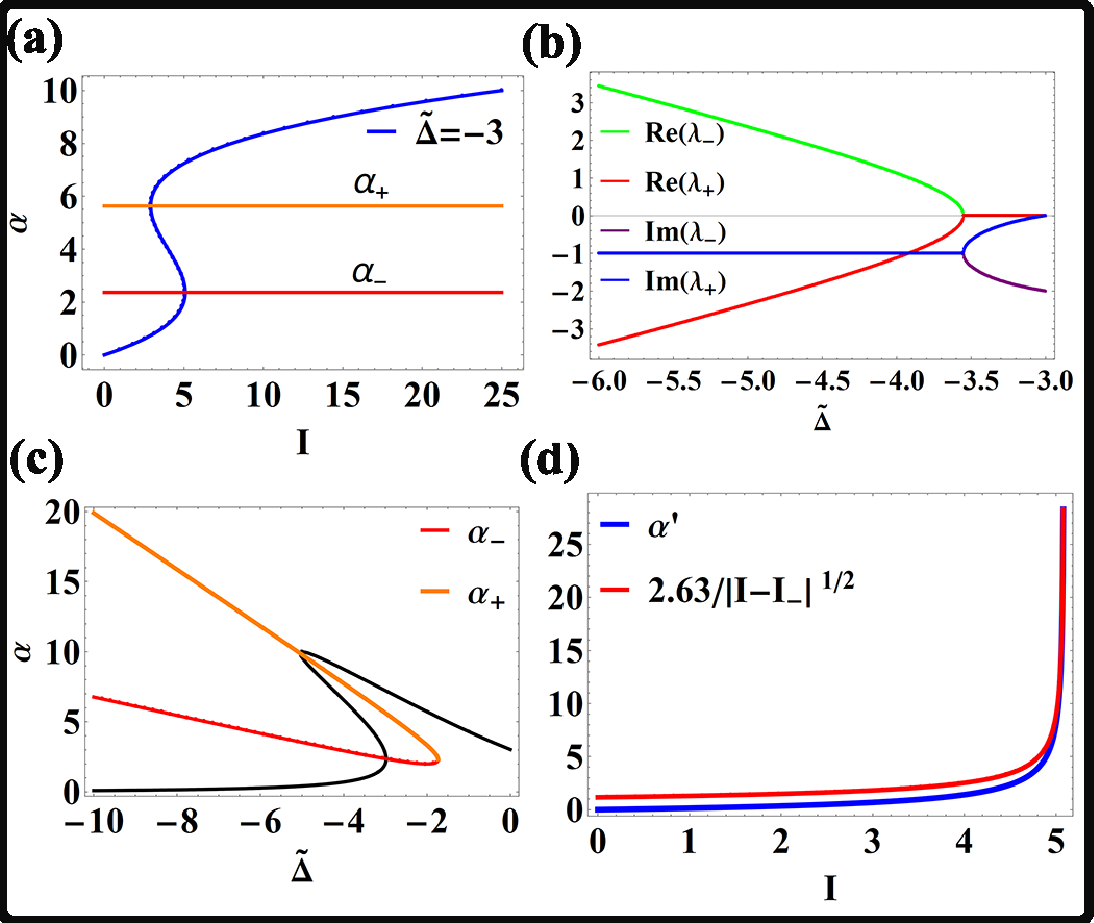}
\caption{(a) The $I$-$\alpha$ curve for the single mode Kerr nonlinear system when $\tilde\Delta=-3$. The turning points are as denoted in the figure; (b) The real and imaginary parts of the eigenvalues of $\mathcal{H}$ as a function of $\tilde{\Delta}$ for $\alpha=2.367$; (c) The $\tilde\Delta$-$\alpha$ curve as $I=5$ (black line) and the turning points $\alpha_\pm$ plotted against $\tilde\Delta$ (orange line and red line, respectively); (d) $\alpha'=\frac{d\alpha}{d\tilde\Delta}$ and $2.63/\abs{I-I_-}^{1/2}$ are plotted against $I$ to make a comparison when $\tilde\Delta=-3$.}
\label{onemode}
\end{figure}

We consider a medium with third order Kerr nonlinearity characterized by a nonlinear contribution to the polarization $P^{(3)}(\omega)=\chi^{(3)}|{E}(\omega)|^{2}{E}(\omega)$ placed in a single-mode cavity with mode variable $a$ as depicted in Fig. \ref{fig_0}. Here, ${E}$ is the cavity electric field and $\chi^{(3)}$ the third order nonlinear susceptibility. The cavity is driven externally at frequency $\omega_d$. The passive nature of nonlinearity indicates that the nonlinear processes are only affected by the frequency composition of the field and not the medium which only plays a catalytic role \cite{lugiato2015nonlinear}. The dynamics of the system in the rotating frame of the drive is given by
\begin{align}
\dot{a}=-i(\Delta-i\gamma)a-i2U|a|^2a+\mathcal{E},
\end{align}
where, $\Delta=\omega_a-\omega_d$, $\omega_a$ denotes the cavity resonance frequency, $U=\frac{3\hbar\omega_a^2\chi^{(3)}}{4\epsilon_0 n V_{eff}}$ is a measure of Kerr nonlinearity of the medium with refractive index $n$, $V_{eff}$ signifies the effective volume of the cavity mode having a leakage rate $\gamma$ and $\mathcal{E}=\sqrt{\frac{2\gamma P_d}{\hbar\omega_d}}$ represents the Rabi frequency of external driving. In the long time limit, the mode $a$ decays into a steady state described by the cubic equation
\begin{align}
I=\frac{\alpha}{2}(1+(\tilde{\Delta}+\frac{\alpha}{2})^2),
\end{align}
where $I=\frac{2U|\mathcal{E}|^2}{\gamma^3}$, $\alpha=4\frac{U}{\gamma}|a_0|^2$, $\tilde{\Delta}=\Delta/\gamma$ and $a_0$ is the steady amplitude of the mode $a$. The Eq. (4) can engender a bistable response under the condition $U\Delta<0$ and $\Delta^2>3\gamma^2$ as illustrated by Fig. \ref{onemode}(a). Notably, there exist two turning points characterized by the coordinates ($I_{\pm}$,$\alpha_\pm$) of the $I$-$\alpha$ curve, subject to $\frac{dI}{d\alpha}=0$, beyond which we observe an abrupt change in $\alpha$. The exact form of $\alpha_\pm$ is given by
\begin{align}
\alpha_\pm=\frac{-4\tilde\Delta\pm2\sqrt{\tilde\Delta^2-3}}{3},
\end{align} 
while $I\pm$ can be obtained from Eq. (4) by substituting the above-mentioned solutions. Moreover, there is a cut off for the pump power beyond which the bistable characteristics set in. The critical magnitude of $I^{c}$ is defined by the inflection point in the $I-\alpha$ graph described by the condition $\frac{dI}{d\alpha}=\frac{d^2I}{d\alpha^2}=0$, providing us
\begin{align}
I^c=-\frac{\alpha^2}{2}(\tilde\Delta+\frac{\alpha}{2}).
\end{align}For a given set of parameters $U$, $\alpha$ and $\gamma$, we would like to perturb the system in $\Delta$, modifying the mode variable into $a=a_0+\delta a$, in which $\delta a$ characterizes the perturbations of the mode $a$ about $a_0$. The dynamics of the perturbations are governed by the following effective Hamiltonian
\begin{align}
\tilde{\mathcal{H}}=\begin{pmatrix}
\tilde{\Delta}+\alpha-i&\beta\\
-\beta^*&-\tilde{\Delta}-\alpha-i
\end{pmatrix},
\end{align}
where $\beta=\frac{2U}{\gamma}a_0^2$. The complex eigenvalues of the Eq. (7) denoted as $\lambda$ refers to the normal modes of the system and they can be obtained by solving the characteristic polynomial equation 
\begin{align}
\lambda^2+2i\lambda+|\beta|^2-(\tilde{\Delta}+\alpha)^2-1=0.
\end{align}
Notably, in the limit when the determinant of the Hamiltonian $(\frac{\alpha}{2})^2-(\tilde{\Delta}+\alpha)^2-1\rightarrow0$, one of the solutions of the Eq. (7) becomes vanishingly small. Note that we are working in the frame rotating at frequency $\omega_d$. Therefore, under this condition, the imaginary part of one of the eigenvalues approaches zero, alluding to the generation of a BIC, as depicted in the Fig. \ref{onemode} (b). It is worth noting that $\alpha\neq0$, i.e., $U\neq0$ is a prerequisite for the existence of such a state. In other words, the generated BIC owes its origin entirely to the Kerr anharmonicities of the mode $a$. For a given value of the parameter $\Delta$, the BICs exist at ($I_\pm,\alpha_\pm$), which are exactly the turning points of the $I-\alpha$ curve as depicted in the Fig. \ref{onemode}(a,c). 

\textit{Application of nonlinearity induced BIC in sensing:}
The existence of BICs also leads to the enhanced sensitivity of the nonlinear response to perturbations in the system parameters. This can be accredited to the existence of the first and second order poles at $\alpha=\alpha_\pm$ in the first order derivative of the nonlinear response 
\begin{align}
\frac{d\alpha}{d\tilde\Delta}=-\frac{8\alpha(\tilde\Delta+\alpha/2)}{3(\alpha-\alpha_-)(\alpha-\alpha_+)},
\end{align}
obtained from differentiating Eq. (4) by $\tilde{\Delta}$. To further elucidate the origin of sensitivity, we expand $I$ around the turning points of the $I-\alpha$ curve, that is, $\alpha=\alpha_\pm$, $I=I_\pm+\frac{\partial I}{\partial \alpha}\epsilon+\frac{\partial^2 I}{\partial \alpha^2}\epsilon^2+O(\epsilon^3)$, where $\epsilon=\alpha-\alpha_{\pm}$ and $I_{\pm}$ are obtained by substituting $\alpha_\pm$ in Eq. (4). Consequently, at the turning points of the curve, $\frac{\partial I}{\partial \alpha}=0$ and we have $\abs{\frac{d\alpha}{d\tilde\Delta}}\sim{{|I-I_\pm|}^{-1/2}}$. On the other hand, close to inflection point sensitivity has the functional dependence $\abs{\frac{d\alpha}{d\tilde\Delta}}\sim{{|I-I_c|}^{-2/3}}$. In practice, one can choose a value of $\Delta$ and the Eq. (8) in conjunction with Eq. (4) respectively determine the corresponding $\alpha_\pm$ and $I_\pm$ appropriate for sensing. As $I$ is varied tantalizingly close to $I_{\pm}$, any perturbations in the parameter $\Delta$ translate into a prodigious shift in the mode response as perceptible from Fig. \ref{onemode}(d). Note that the sensitivity to aberrations in $\Delta$ is a direct artifact of the existence of a BIC. 

Bearing in mind the generality of our analysis, it is interesting to observe the variety of experimental platforms available to implement our scheme for investigating BICs produced by nonlinearity induced extra dimensions. Some of the well-known examples in the context of passive nonlinearities and bistability include Sodium vapor \cite{PhysRevLett.36.1135}, Ruby \cite{venkatesan1977optical}, Kerr liquids like $CS_2$, nitrobenzene, electronic nonlinearity of Rb vapor etc. to name a few \cite{gibbs2012optical, boyd2020nonlinear}. In the subsequent section, we stretch the analysis into the case of a two mode anharmonic system.

\section{Engineering BIC in a two mode Kerr nonlinear system}\label{sec3}
We begin this section by considering a two mode active Kerr nonlinear system that consists of modes $a$ and $b$ coupled coherently through a real parameter $g$, and $b$ is externally pumped at a frequency of $\omega_d$. The Hamiltonian of the system can be expressed as 
\begin{equation}
\begin{aligned}
H/\hbar&=\omega_aa^\dagger a+\omega_b b^\dagger b+g\left(b^\dagger a+ba^\dagger\right)\\
&\!\!\!\!\!\!+Ub^\dagger bb^\dagger b+i\Omega\left(b^\dagger e^{-i\omega_d t}-be^{i\omega_d t}\right),
\end{aligned}
\label{equ:cm}
\end{equation}
where $\omega_a$ and $\omega_b$ represent the resonance frequencies of the modes $a$ and $b$, the coefficient $U$ quantifies the strength of Kerr nonlinearity, and $\Omega$ denotes the Rabi frequency of external driving. The systems characterized by the aforementioned Hamiltonian are prevalent in nature, for example, a collection of two-level atoms under the conditions of no saturation which act as an active Kerr nonlinear medium in a driven resonant cavity. The dynamics of the system in the rotating frame of the drive is provided by
\begin{equation}
	\begin{aligned}
		&\dot a=-\left(i\delta_a+\gamma_a\right)a-igb,\\
		&\dot b=-\left(i\delta_b+\gamma_b\right)b-2iUb^\dagger bb-iga+\Omega,
	\end{aligned}
\end{equation}
where $\delta_a=\omega_a-\omega_d$, $\delta_b=\omega_b+U-\omega_d$, and $\gamma_a$ and $\gamma_b$ denote the dissipation rates of the modes $a$ and $b$, respectively. In the long-time limit, the system decay into a steady state, i.e., $a\rightarrow a_0$, $b\rightarrow b_0$ lending the following nonlinear cubic equation 
\begin{equation}
	I=4 x^3+4\tilde\delta_{R}x^2+\left|\tilde\delta\right|^2x,
	\label{equ:Ix}
\end{equation}
where $I=U\Omega^2$, $x=U\left|b_0\right|^2$, $\tilde\delta=\delta_b-i\gamma_b-\frac{g^2}{\delta_a-i\gamma_a}$, and we define $\tilde\delta_R=\delta_b-\frac{ g^2\delta_a}{\delta_a^2+\gamma_a^2}$ and $\tilde\delta_I=-\gamma_b-\frac{g^2\gamma_a}{\delta_a^2+\gamma_a^2}$ as the real and imaginary parts of $\tilde\delta$, respectively. Notice that $\tilde\delta_I$ is negative. Under the criterion $\tilde \delta_{R}<\sqrt3\tilde\delta_{I}$, there exist three possible roots for $x$, leading to a bistable response, wherein, two of the roots are stable while the third is unstable. 

\textit{Conditions for the existence of BIC:}
To analyze the effect of perturbations around the steady state, we use a linearized approximation by letting $a=a_0+\mathscr A$ and $b=b_0+\mathscr B$, where $\mathscr A$ and $\mathscr B$ signify the perturbations of mode $a$ about $a_0$ and mode $b$ about $b_0$, respectively. The dynamics of the perutrbations $\psi^T = \left[\mathscr A,\mathscr B,\mathscr A^\dagger,\mathscr B^\dagger\right]$ are governed by the following equation,
\begin{equation}
	\frac{\partial\psi}{\partial t}=-i\mathcal H\psi+\mathcal I,
\end{equation}
where $\mathcal H$ is the effective Hamiltonian
\begin{equation}
	\mathcal H=\left(
	\begin{array}{cccc}
		\delta_a-i\gamma_a & g & 0 & 0\\
		g & \delta_b+4x-i\gamma_b & 0 & 2Ub_0^2\\
		0 & 0 & -\delta_a-i\gamma_a & -g\\
		0 & -2Ub_0^{*2} & -g & -\delta_b+4x-i\gamma_b\\
	\end{array}
	\right),
	\label{equ:ha}
\end{equation}
and $\mathcal I=0$ for the steady state. The normal modes of the system are hallmarked by complex eigenvalues of Eq. (\ref{equ:ha}), which can be obtained by solving the characteristic polynomial equation $\det\left(\mathcal H-\lambda\mathbf I\right)=0$. Conspicuously, when $\det \mathcal{H}=0$, one of the eigenvalues can approach zero (in the rotating frame of the drive), spawning real eigenvalues and thereby indicating the emergence of a BIC. Therefore, we first determine the parameter domain consistent with condition 
\begin{equation}
	\begin{aligned}
	0&=\det\mathcal H\\
	&=12\left(\delta_a^2+\gamma_a^2\right)x^2+8\left(-\delta_a g^2+\delta_b\delta_a^2+\delta_b\gamma_a^2\right)x\\
	&\ \ \ \ \ \ +\left(g^2-\delta_a\delta_b+\gamma_a\gamma_b\right)^2+\left(\delta_a\gamma_b+\delta_b\gamma_a\right)^2.
	\end{aligned}
\label{equ:det}
\end{equation}
It is worth noting that the existence of BIC relies on the prerequisite $x=U\left|b_0\right|^2\neq0$. In other words, the Kerr anharmonicities of the mode $b$ are solely responsible for the creation of the BIC. Upon solving Eq. (\ref{equ:det}), we discover that BICs can exist at points
\begin{equation}
	x_{\pm}=-\frac13\tilde \delta_{R}\pm\frac16\sqrt{\tilde \delta_{R}^2-3\tilde\delta_{I}^2},
\end{equation}
which are exactly the turning points of the $I$ -- $x$ curve given in Eq. (\ref{equ:Ix}), obtained from solving the condition $\frac{dI}{dx}=0$.

While invoking the linearized dynamics, one must make sure that the dynamical system is stable, which is to ensure that the eigenvalues of $\mathcal{H}$ have negative imaginary parts. Consequently, we define $\lambda_R$ and $\lambda_I$ as the real and imaginary parts of the complex eigenvalues, respectively, and let $\lambda'=-i\lambda$. The characteristic polynomial equation can then be written as
\begin{equation}
	0=\det\left(\mathcal H-i\lambda'\mathbf I\right)=\lambda'^4+a_1\lambda'^3+a_2\lambda'^2+a_3\lambda'+a_4,
\end{equation}
where
\begin{equation}
	\begin{aligned}
		&a_1=2\left(\gamma_a+\gamma_b\right),\\
		&a_2=\delta_a^2+2g^2+\left(\gamma_a^2+4\gamma_a\gamma_b+\gamma_b^2\right)+\left(12 x^2+8\delta_b x+\delta_b^2\right),\\
		&\begin{aligned}a_3&=2\delta_a^2\gamma_b+2\delta_b^2\gamma_a+2\left(\gamma_a\gamma_b+g^2\right)\left(\gamma_a+\gamma_b\right)\\
			&\ \ \ \ \ \ +16\delta_b\gamma_ax+24\gamma_a x^2,
		\end{aligned}\\
	    &a_4=\det \mathcal H.
	\end{aligned}
\end{equation}
The stability conditions of the system can be obtained by employing the Routh-Hurwitz Criteria, yielding the constraints $a_1>0$, $a_3>0$, $a_4>0$, and $a_1a_2a_3>a_3^2+a_1^2a_4$. Apparently, the first two conditions are met automatically, and we find
\begin{equation}
	\begin{aligned}
		&a_1a_2a_3-a_3^2-a_1^2 a_4=4\gamma_a\gamma_b\left(12 x^2+8\delta_b x-\delta_a^2+\delta_b^2\right)^2\\
		&\!\!\!\!\!\!+4\gamma_a\gamma_b\left(\gamma_a+\gamma_b\right)^2\left[24x^2+16\delta_b x+2\left(\delta_a^2+\delta_b^2\right)+\left(\gamma_a+\gamma_b\right)^2\right]\\
		&\!\!\!\!\!\!+4g^2\left(\gamma_a+\gamma_b\right)^2\left[12x^2+8\left(\delta_a+\delta_b\right)x+\left(\delta_a+\delta_b\right)^2+\left(\gamma_a+\gamma_b\right)^2\right],
	\end{aligned}
\end{equation}
which is manifestly positive fulfilling the final criterion. The only remaining criterion $a_4=\det\mathcal H>0$ is satisfied along with $\tilde \delta_{R}<\sqrt3\tilde\delta_{I}$ and $x\in (0,x_-)\cup(x_+,\infty)$.

\textit{Sensing capabilities of nonlinearity induced BIC:}
The importance of the above results can be legitimized in the optical domain with several well known systems, including, for instance Sagnac resonators \cite{zhang2020breaking, silver2021nonlinear} among other settings \cite{boyd2020nonlinear}. 
The presence of BICs at points $x_\pm$ contributes to the significantly improved sensitivity of the nonlinear response to variations in the system parameters, in particular, to perturbations in natural frequency of the active nonlinear medium. The remarkable sensitivity is a direct upshot of the existence of first or second order poles at $x=x_\pm$ in the first derivative of the nonlinear response which has the functional form 
\begin{equation}
	\frac{d x}{d\delta_b}=-\frac{x(x+\tilde\delta_R/2)}{3(x-x_-)(x-x_+)},
\end{equation}
analogous to Eq. (9). Therefore, it immediately follows that adjacent to the turning points, we have $\abs{\frac{d x}{d \delta_b}}\sim\abs{{I(x_\pm)-I}}^{-1/2}$. By the same token, close to the inflection point, the sensitivity scales as $\abs{I_c-I}^{-2/3}$. 

\begin{figure}
 \captionsetup{justification=raggedright,singlelinecheck=false}
 \centering
   \includegraphics[scale=0.46]{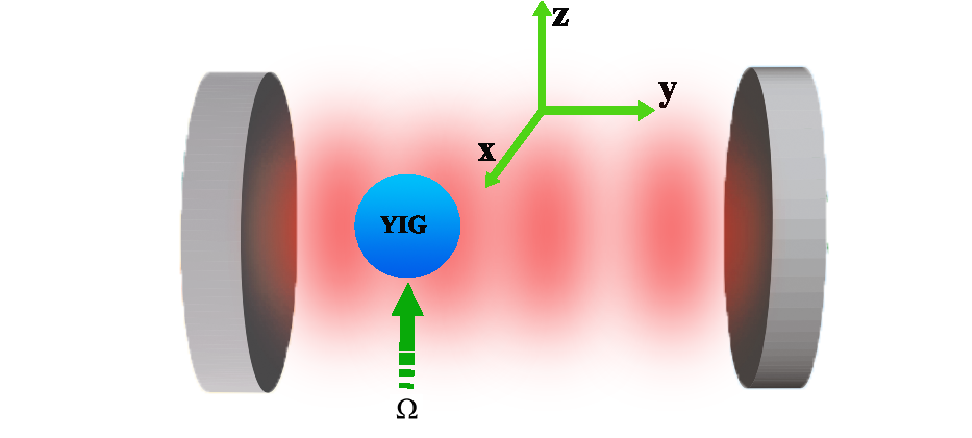}
\caption{Schematic of the cavity-magnon system}
\label{sch}
\end{figure}

\begin{figure}
 \captionsetup{justification=raggedright,singlelinecheck=false}
 \centering
   \includegraphics[scale=0.46]{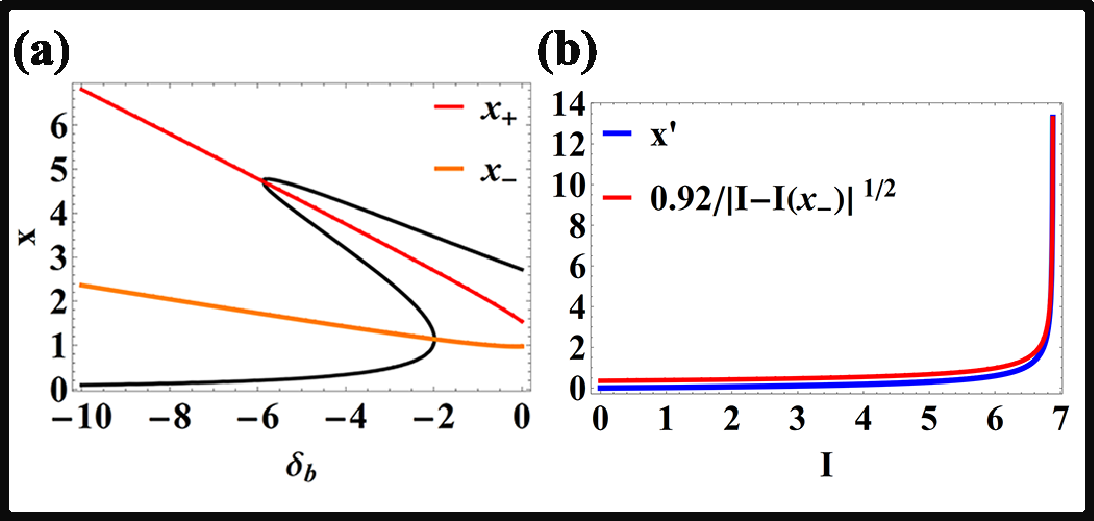}
\caption{(a) The $\delta_b$-$x$ curve as $I=18$ (black line) and the turning points $x_\pm$ plotted against $\delta_b$ (red line and orange line, respectively). (b) $x'=\frac{dx}{d\delta_b}$ and $0.92/\abs{I-I(x_-)}^{1/2}$ are plotted against $I$ to make a comparison when $\delta_b=0$. Parameters are $\gamma_a=\gamma_b=1$, $g=\delta_a=4$.}
\label{twomode}
\end{figure}

\textit{Sensing in magnetic systems:} In view of the extensive studies on nonlinearities \cite{PhysRevB.94.224410} in ferrimagnetic spheres, it is worthwhile to consider magnetic systems to implement the sensing scheme. Note that the anharmonicities in optical systems is a direct consequence of the nonlinear response of electrical polarization. In stark contrast, the anharmonic component in a magnetic system originates from the nonlinear magnetization. We consider a single ferromagnetic YIG interacting with a microwave cavity as portrayed in Fig. \ref{sch}. The ferromagnet couples strongly with the microwave field at room temperature, giving rise to quasiparticles, namely cavity-magnon polaritons. The YIG acts as an active Kerr medium, which can be pinned down to the magnetocrystalline anisotropy \cite{PhysRevB.94.224410, PhysRevLett.127.183202, PhysRevLett.120.057202} of the sample. A strong microwave pump of power $P_d$ and frequency $\omega_d$ is used to stimulate the weak anharmonicity of the YIG, which is of the order $10^{-9}$ Hz. The full Hamiltonian of the cavity-magnon system is consistent with Eq. (\ref{equ:cm}) where the mode operators $a$, $b$ are respectively superseded by cavity and magnon annihilation operators. The quantities $\omega_a$ and $\omega_b$ represent the cavity and Kittel mode resonance frequencies. Rabi frequency of external pumping takes the form $\Omega=\gamma_e\sqrt{\frac{5\pi\rho d P_d}{3c}}$, where $\gamma_e$ is the gyromagnetic ratio, $\rho$ denotes the spin density of the YIG with a diameter $d$ and $c$ stands for the velocity of light. For experimentally realizable parameters of the system, we plot in Fig. \ref{twomode} $x$ from Eq. (12) by varying $\delta_b$ and $I$ and the results replicate the physics described in Fig. \ref{onemode}.
\section{Conclusions}\label{sec4}
In conclusion, we have demonstrated a new scheme apropos of single and two-mode Kerr nonlinear systems to engineer BICs. In the context of single mode systems, we considered a passive Kerr nonlinearity in an optical cavity that demonstrates bistability. As the the system parameters are tuned in close proximity to the turning points of the hysteresis, a BIC springs into existence marked by a vanishing linewidth of the mode. In the neighborhood of the BIC, the steady state response was observed to show pronounced sensitivity to perturbations in the detunings. This remarkable sensitivity can be traced down to the existence of poles in the first order derivative of the response with respect to the perturbation variable. The sensitivity to perturbations scales as inverse square root of the deviations in external pump powers optimal for the turning points. Further, we extended the analysis into the regime of two-mode systems possessing an active nonlinear medium. Our analysis is generic, applicable to a large class of systems, including, both optical and magnetic systems. Some of the passive nonlinear optical platforms include nonlinear media like $CS_2$, nitrobenzene, Rb vapor whereas high-quality Sagnac resonators support active Kerr nonlinearities. In addition, we considered an active Kerr medium provided by magnetic systems interacting with a microwave cavity where research activity has flourished of late. In the domain of large detunings of the active Kerr medium, the two-mode setup can be described by an effectively single-mode anharmonic system in lockstep with the results from the passive Kerr nonlinearity in an optical cavity. 
\section{Acknowledgements}
The authors acknowledge the support of The Air Force Office of Scientific Research [AFOSR award no FA9550-20-1-0366], The Robert A. Welch Foundation [grant no A-1943] and the Herman F. Heep and Minnie Belle Heep Texas A\&M University endowed fund.

Q. M. and J. M. P. N. contributed equally to this work. 
\appendix
\section{Equivalence between single-mode and two-mode anharmonic system}
\label{Appendix:a}
So far, we have discussed schemes for the creation of BICs in single and two mode nonlinear systems. It is worth mentioning that there exists a close correspondence between the two mode and single mode results in the limit of large $\delta_b$. To enunciate this, let us delve into the second part of Eq. (11). In the long-time limit, we have
\begin{align}
-\left(i\delta_b+\gamma_b\right)m-2iUb^\dagger bb-iga+\Omega=0.
\end{align}
Note that the effect of $\gamma_b$ pales in comparison with $\delta_b$ and we can recast the above equation into
\begin{align}
 b=-\frac{(ga+i\Omega)}{\delta_b}\Big[1+x\Big]^{-1},
\end{align}
where $x=\frac{2U|b|^2}{\delta_b}$. For the purpose of simplification, we set $\Omega=0$ and assume that the external drive is on the cavity at Rabi frequency $\mathcal{E}$. Owing to the largeness of $\delta_b$, it is discernible that $x<<1$. Therefore, we can revise the above equation as
\begin{align}
b=-\frac{(ga+i\Omega)}{\delta_b}\Big[1-x+O(x^2)\Big].
\end{align}
Keeping only terms up to first order in $x$, we are left with $b=-\frac{ga}{\delta_b}\Big[1-\frac{2U|b|^2}{\delta_b}\Big]$. Upon iterating the solution and omitting the higher order terms, the approximate solution for $b$ morph into
\begin{align}
b=-\frac{ga}{\delta_b}+2\Big(\frac{g}{\delta_b}\Big)^3\Big(\frac{U}{\delta_b}\Big)|a|^2a.
\end{align}
Substituting this into the first part of Eq. (11), we obtain an effective single mode description of the dynamics of the system,
\begin{align}
\dot{a}=-(i\tilde{\delta}_a+\gamma_a)a-i\tilde{U}|a|^2a+\mathcal{E},
\end{align}
where $\tilde{\delta}_a=\delta_a-\frac{g^2}{\delta_a}$ and $\tilde{U}=2\Big(\frac{g}{\delta_b}\Big)^4U$. Strikingly, the preceding equation reproduces Eq. (3) with $\Delta$, $\gamma$ and $U$ respectively replaced by $\tilde{\delta}_a$, $\gamma_a$ and $\tilde{U}$, unfolding the equivalence between two-mode and single-mode nonlinear systems in the realm of large $\delta_b$.  
\bibliography{references}
\end{document}